# Drowsiness detection using combined neuroimaging: Overview and Challenges

A S M Sharifuzzaman Sagar*, Tajken Salehen, Md Abdur Rob

*Intelligent Mechatronics Engineering, Sejong University, Seoul, South Korea*
*American International University of Bangladesh, Dhaka, Bangladesh*
*School of Mechatronics Engineering, China University of Mining and Technology, Xuzhou, China*

**Abstract**

   Brain-computer interfaces (BCIs) collect, analyze, and convert brain activity into instructions and send it to the detection system. BCI is becoming popular in under-brain activities in certain conditions such as attention-based tasks. Researchers have recently used combined neuroimaging techniques such as EEG+fNIRS and EEG+fMRI to solve many real-world problems. Drowsiness detection or sleep inertia is one of the central research areas for the combined neuroimaging techniques. This paper aims to investigate the recent application of combined neuroimaging-based BCI on drowsiness detection or sleep inertia. To this end, this is the only overview paper of the combined neuroimaging-based drowsiness detection system.

## 1. Introduction

   Brain-computer interaction (BCI) is a method of recording neurological phenomena of the human brain and helps interact with the surroundings without the help of any muscular movement of the other part of the body [1]. The communication between brain signals and external devices that interpret the signals can be wired or wirelessly. The BCI was first introduced for biomedical applications; later, it was used in numerous assistive devices [2]. The BCI, over the years, has facilitated restoring movement or physical activity of disabled persons. Brain-computer interaction. The BCI can be divided into active BCI and passive BCI. Active BCI system mainly consists of intentional motor imagery and external stimulation of P3000[4-6]. Passive BCI interprets unintended cognitive states of the human brain. EEG is the most common technology used in BCI; in addition fNIRS, fMRI, EEG are also used BCI systems [7—14].

*Corresponding author
   E-mail addresses: sharifsagar80@sju.ac.kr, tajkensalehen@gmail.com, arob2664@gmail.com

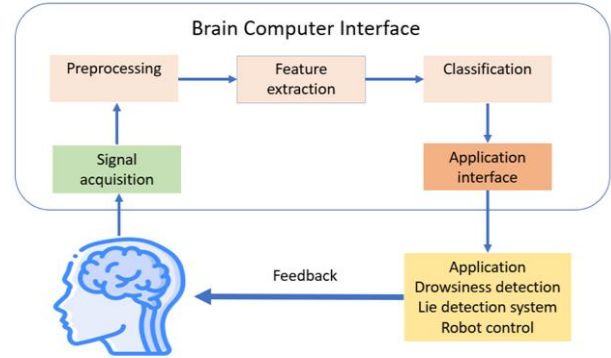

Figure 1. Basic Architecture of brain-computer interaction.

   Fig. 1. shows the basic structure of BCI. The BCI mainly consists of three parts such as signal acquisition, feature extraction, and feature classification [15]. Signal acquisition parts record the brain signal through hardware. Then the signals go to noise filtering processing as brainwaves are not stable and produce noises [16]. The feature extraction parts identify essential data from the signal and send them to feature classification algorithms for further processing [17]. The feature classification interprets data fed by the feature extraction to useable information for pattern recognition.

*2.1 Drowsiness detection*

   Drowsiness detection technology is one of the most popular use cases of BCI. Drowsiness detection technology is mainly used to prevent drivers' workers from falling asleep during their relevant work. Most research focuses on detecting driver drowsiness detection, one of the biggest causes of road accidents. Drowsy driving means when the driver becomes exhausted or sleepy while driving on roads, making it very difficult to focus on driving. Drowsy driving occurs when drivers do not get adequate sleep or work as commercial drivers for long shifts or suffer from sleep disorders, and more [18]. Therefore, the driver leans towards a slight perceptive loss, which results in a slow reaction movement. The driver may fall asleep while driving the vehicle [19].

Fig. 2. shows the general framework of the BCI-based drowsiness detection system. BCI-based drowsiness detection detects through analyzing neuroimages of the brain. Different approaches such as EEG, fNIRS, fMRI are being used as the input of the drowsiness detection system. Then they undergo preprocessing. After preprocessing, the system extracts the feature and classifies them using different SVM, KNN, and deep learning techniques to detect a person's drowsiness.

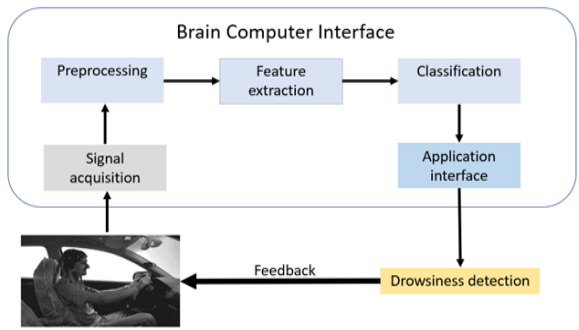

Figure 2 BCI-based Drowsiness detection system.

## 2. Combined Neuroimaging based Drowsiness detection

Recently, researchers have been exploring combining different approaches of BCI to solve a real-world problem. The most common approaches of Combined BCI include fNIRS + EEG, EEG+fMRI because these approaches have the potential to interpret brain signals accurately for specific problems such as lie detection, drowsiness detection, sleep pattern analysis. This report discusses the two most used approaches for analyzing drowsiness detection or sleep/wake-up analysis, such as fNIRS + EEG, EEG+fMRI.

*2.1 fNIRS+EEG based drowsiness detection*

Nowadays, Researchers are investigating the integration of EEG and fNIRS approaches to detect drowsiness detection. This method includes acquiring EEG and fNIRS signals, preprocessing the data, feature extraction of the acquired signal, and integrating them into one classifier to detect or interpret the signals to meaningful solutions.

Dehais et al. proposed a PassiveBCI based drowsiness detection system to monitor the drowsiness of plane pilots using the combination of EEG and fNIRS signal [20]. The authors recruited four pilots from a pilot training school to experiment with flight simulators (*ISAE-SUPAERO*) and real planes (DR400). These traffic patterns were divided into five actions pilots need to perform: the take-off leg, the crosswind leg, the downwind leg, the base leg, and the final landing. The authors used a Neuroelectrics system to acquire the brain signal from the 23 channels at 500Hz, and the remaining channels were removed to facilitate the fNIRS signal acquisition. The authors used the NIRSport NIRX sensor to record the fNIRS signal at 8.93Hz. The acquired data were synchronized and analyzed by Matlab R2015b and Homer2 software package [21]. The acquired data were spanned into successive and non-overlapping 1-minute time windows before analysis. The authors used Automatic Subspace Reconstruction to remove Inconsistent signals from the EEG data to acquire smooth data segments. On the other hand, fNIRS signals were converted to optical density to remove the high deviance parts of the signal. Then the low pass filter of order 2 (0.5 Hz) and high pass filter of order 5 (0.01Hz) was used to smooth the signal. The authors used shrinkage linear discriminant analysis (sLDA) on the extracted features to classify the EEG signals. The authors have achieved an accuracy of 87.2% for simulation software and an accuracy of 87.6% for an actual flight condition when combining EEG and fNIRS features.

Chuang et al. investigate a drowsy driver's brain dynamics to avoid potential accidents [22]. The authors particularly observe the brains' electrodynamics and hemodynamics during the driving condition. The author recorded EEG and fNIRS signals simultaneously from 16 subjects. A virtual driving simulator having six degrees of freedom and actual car equipment was selected for the experiment. The experiments were conducted for one hour, where the subject was asked to drive the lane most of the time and change lanes throughout 1 hour. The EEG and fNIRS signals were recorded during the lane departing by keeping the minds of three events: deviation onset, response onset, and response offset. Synamp2 system with a 32-channel electrode was used to capture the EEG data. The EEG signals were amplified and recorded at a sampling rate of 1,000 Hz using a 32-bit Analog to a Digital Converter. The NIRScout system was used to acquire the fNIRS signal. It has 8 sources and 15 detectors that can record fNIRS signals at a rate of 7.81Hz. The fNIRS signal from the right and left hemispheres were collected through 18 channels of the NIRscout system. The authors used the EEGLAb toolbox from MATLAB to analyze the EEG data and they used the nirsLAB toolbox to analyze the fNIRS data. They used a high pass filter and a low pass finite impulse response filter to eliminate noises from the signal. Then, the authors used Fast Fourier transform to convert the data into the frequency domain and divided it into four frequency bands: Delta, Theta, Alpha, and Beta. Additionally, the authors used SPSS software to calculate the Pearson correlation between EEG frequencies and behavioral performance. With the rise in RT, the authors noted an enhanced occipital pre-stimulus strength in theta and alpha bands. The RT- and the EEG-Record-Power pre-stimulus correlation coefficients were Pairson's r=0.88 and Pearson's r=0.91. They also reported a considerable rise in HbO power throughout the deviation period in comparison with the baseline (p-value <0,05) for the fNIRS signal. This increased HbO power is similar to the increased EEG alpha power in the occipital area. The authors concluded that subjects were in fatigue conditions during driving experiments.

Lin et al. also explore the physiological phenomena of a drowsy driver using simultaneous fNIRS and EEG signals to understand the relationship between hemodynamics and electrical features and driving performance [24]. The authors selected 16 healthy subjects having good vision and license for the experiment. The subjects were asked to participate in a virtual simulation with four endless lane roads. Data gathered throughout the experiment were captured concurrently with the behavioral, EEG, and fNIRS. RTs were continually collected with the WTK software, and the EEG data were recorded at 2000Hz using a V-Amp 16 expert system with 128 channels in a conventional NIRS cap placement as per the standard 10-20 system. The authors used EEGLAB from the MATLAB toolbox with a bandpass filter of between 1 to 50Hz to analyze the signal. Later, the signals were downsampled to 250Hz for better data representation and converted to the frequency domain using fast Fourier transform. The authors divided the EEG signal into four frequency bands: delta(1-3Hz), theta(4-7Hz), alpha(8-12Hz), beta(13-30Hz). The authors used "modified Beer-Lambert law" from the nirsLAB toolbox to convert NIRS data from light absorption to relative concentrations of HbO2 and HbR (unit: mM). They also used high-pass(0.015Hz) and low-pass filtering (0.08Hz) of HbO2 and HbR to avoid possible noise from the heartbeat, breathing, and low-frequency signals. After preprocessing, the authors determine the driver's fatigue level based on the reaction time of the drivers during lane deviation events. Based on the reaction time, the authors determined three performance groups: optimal performance (reaction time radio ≤1), suboptimal performance (reaction time ratio between 1 to 2), and poor performance (reaction time ratio ≥2).

The authors observe two variations during experiments, one before the deviation task (tonic variations) and another (phasic variations) afterward the deviation. According to the authors, tonic data indicate a greater percentage of oxygenated hemoglobin (HbO2) and EEG theta, alpha, and beta changes. These dynamics have a significant correlation with sleepy driving. The phase EEG results indicate that the steering vehicle is desynchronized due to an event in all power bands. As behavior worsened, the phasic HbO2 content varied. Furthermore, the negative correlations between static EEG delta and alpha power and HbO2 oscillation indicate that HbO2 activation is associated with sleepiness.

Ahn et al. examine the neuro-physical relation of driver fatigue through simultaneous EEG and fNIRS data [25]. The authors extracted the potential features of each modality to determine sleep-deprived conditions. Lastly, the authors proposed a driving condition level (DCL) to determine whether a driver is drowsy or not. The authors selected eleven healthy subjects having valid driver's licenses for the experiments. They were asked to perform a virtual driving simulation task. All subjects performed the simulation task under well-rested and sleep-deprived conditions. The authors placed 64 electrodes on the driver's scalp based on the 10-20 international system to acquire EEG signals. EEG data were acquired at a sampling rate of 512Hz through BCI2000 software. The author also used a custom-built fNIRS system to acquire hemodynamic changes in the brain at a rate of 10Hz. They used EEGLAB from the MATLAB software to calculate power spectral density and relative power level (RPL) to reduce session variability. The authors looked at the RPL values and discovered that the RPL values for delta, theta, and gamma were not significantly different between the two driving scenarios. The authors also observed a clear feature distinction of EEG and fNIRS signal between well-rested and sleep-deprived tests. The authors used those features to construct DCL to determine the neuro-psychological correlation of the drowsy driving scenario. Then the authors used the DCL approach to classify the drowsiness state of a driver and achieved a mean accuracy of 68.3 for the EEG+fNIRS approach.

Ngyuen et al. presented a combined EEG and NIRS method to driver fatigue predictions [26]. The authors recruited eleven health subjects and were asked to take enough sleep before participation day. The experiment was done using a driving simulator system which consists of a brake, wheel, chair. A camera was also installed to monitor the participant's behavior. EEG signal was acquired by Biosemi Active Two system at a sampled rate of 512Hz. On the other hand, The NIRS data were collected with though an 8 channel NIRS system at a 10Hz rate.

The authors used a 60 Hz notch filter to remove power line noise from the captured signal. They also used bandpass filters (1-50 Hz) for EEG data and low-pass filters (0.2 Hz) for NRIS data in order to smooth the data for further processing. The EEG signal was further broken down into 64 ICA components to eliminate bad components prior to feature extraction. To minimize the variability across individuals, RPL was computed by distributing band power by the total of the output. The hemodynamics responses from the NIRS modality used in the experiment were the HbO, Hb, and THb differences. The HbO and Hb changes were calculated directly from the measured light intensity, whereas the THb variation was calculated by adding the HbO and Hb variations together. Then the authors FLDA based classifier to generate classification accuracy. A group of outputs with regard to the classifier was undertaken to explore the overall impact of the classification utilizing EEG and fNIRS data. The authors achieved a mean accuracy of 79.2%, which outperforms the accuracy of classifying drowsiness by using EEG and NIRS signals alone.

*2.1 fMRI+EEG based drowsiness detection*

Researchers are investigating the integration of EEG and fMRI approaches to solving real-world problems. However, Combined neuroimaging techniques such as EEG+fMRI were not implemented directly to detect drowsiness; some works related to drowsiness, such as sleep inertia analysis and sleep-wake state analysis, were done using EEG+fMRI BCI systems.

Vallat et al. examined changes in behavior execution, EEG spectral power, and fMRI functional resting-state connection over three data acquisition times: before an early night's sleep after a nap awakes and after waking [27]. The authors chose 55 research participants based on their reports with a regular sleep-waking pattern, small sleep disorders, and an MRI brain scan in advance. The authors recorded polysomnography using a 15-channel MR-compatible sleep study cap which included 9 EEG electrodes. The semi-automatic MRI artifact rejection methods developed in Brain Vision Analyzer 2.1 software were used to eliminate gradient switching and cardiac pulse (cardioballistic effect) artifacts in each of the signal segments. The data were normalized to 500 Hz and bandpass filtered between 0.1 and 40 Hz. Power spectral density (PSD) was calculated for four frequency bands delta, theta, alpha, beta using the MNE-python packageICA. Moreover, the authors acquired the MRI images from a Prisma 3T scanner at the Primage neuroimaging center. The authors calculated the interplay among various response variables, including DST efficiency, EEG spectral measurements, sleep characteristics, and the average connection between and within brain networks. The authors discovered that cognitive performances decrease during the first few minutes after waking up from sleep, and sleep-specific activity incursion reflects an increased spectral strength of slow activity and a primary loss in brain network function.

Czisch et al. investigated the effects of 36-hour complete depletion of sleep using EEG and fMRI [29]. In particular, they concentrated on alterations in the attention developments caused by the dynamic acoustic oddballs' task, which are capable of isolating trials with objective EEG indications of high alertness corresponding to warning conditions and decreased vigilance corresponding to sleepy conditions. The authors selected 20 healthy subjects to perform an oddball acoustic task having a well-rested night and after 36h of total sleep deprivation. In addition to 11 EEG channels, an MR-compatible EEG device was used to monitor electrophysiological parameters. The EEG signals were processed at a frequency range of 0.5 to 70 hertz. The authors used the conventional visual scoring technique to evaluate the EEG data from all runs, which was followed by the sub-classification by Valley technique [31]. This gives the further definition of drowsiness: Stages 1A and 1B of the transition from wakefulness to sleep are distinguished by the presence of slower awake alpha rhythm intertwined with intermediate power frequency structure, as well as partial or measurable slow rolling eye movements. Moreover, they used an eight-channel angiography and a 1.5-T scanner to acquire functional magnetic resonance imaging (fMRI) data. Images were analyzed using MATLAB 2008b. The authors noted that activity relating to oddball tasks seems to be supported by adaptive co-activation of insular areas in the alert condition, although the task-negative activity in the network's nonattendance state changes after total sleep deprivation. On the other hand, positive activities at the oddball task were seriously reduced under sleepy or drowsy conditions, but following a well-rested night, activity at the task-negative showed levels similar to the control condition.

Chen et al. explore the neurophysiological processes behind sleep inertia through a correlation between before and after sleep and eye-open rest with EEG-fMRI [32]. For the experiments, the authors chose 55 people between the ages of 18 and 25. The authors design three recording phases for the experiments: pre-sleep, sleeping, and sleep inertia. For each of the three stages, simultaneous EEG–fMRI recordings were carried out on the participants. Pre-sleep and sleep inertia periods were conducted with individuals looking at a fixation point on a screen while relaxing and being instructed not to think about anything consciously throughout the study. The authors used a 32-channel Brain Products-compatible brain device to acquire EEG signals. According to the 2017 AASM handbook, the authors evaluated the EEG signal using a 30-second time frame [33]. The EEG signals were then sampled to 250 Hz and filtered using a ChebyshevII-type filter, utilizing a frequency range of 0.1–45 Hz. The processed EEG data were segmented into three-second segments, corresponding to two fMRI scans. The power for the three frequency bands: delta, theta, and alpha, was determined using FFT. The GIFT toolkit was used to extract resting-state networks using group ICA. The optimum number of components was determined using the minimum description length criteria set at 25. Principal component analysis (PCA) was used to reduce data prior to ICA. The Extended Infomax method was used to perform ICA decomposition analysis on concatenated datasets.

The authors performed a temporal correlation study between the EEG observance and the time courses of fMRI. The authors discovered a significant correlation value ("r = 0.329, p =.001") in the pre-sleep phase, which vanished during sleep inertia ("r = 0.037, p =.602"). The authors observed the sleepy features of decreasing EEG power and decreased BOLD function while a person is still in sleep inertia. They also discovered that a person with greater EEG alertness exhibited greater front-parietal network (FPN) activity; however, this characteristic vanished during sleep inertia.

Lei et al. examined simultaneous EEG-fMRI data obtained from the healthy individuals throughout rest awareness and non-rapid eye movement sleep (NREM) [34]. The authors selected 36 participants for the experiment. The authors used a non-magnetic MRI-compatible EEG device to digitize the EEG signal at 5 kHz and reference it online to FCz. Each of the 32 electrodes was a ring-type sintered non-magnetic Ag/AgCl electrode, which was put on the scalp using the international 10/20 method. Moreover, the authors used a 3T Siemens Trio scanner to obtain a high-resolution T1-weighted structural volume. The T1-weighted structural volume at high resolution served as an anatomical standard for the functioning scan. 500 functional volumes were scanned with fMRI using the EPI sequence. They used the EEGLAB FMRIB toolbox to rectify the MRI imagery artifact offline and down-sampled the data at

250Hz. Temporal ICA was used to reduce the BCG, ocular, and residual image artifacts in the EEG data. The recordings were split into 30-second blocks labeled as alertness, rapid eye movement, and NREM sleep. The authors used a 2-factor analysis of variance (ANOVA) to perform statistical analysis for both the scaling factor of EEG and fMRI in various individuals. The authors utilized post-hoc t-tests to detect major changes from sleep to rest. After corrections for multiple comparisons using the Bonferroni technique, differences were estimated significantly when a probability was below 0.05. They also utilized the Spearman correlation for EEG and fMRI data to measure the match between medium power and the slope. The authors computed and averaged the value for the various EEG electrodes since the scaling exponents from EEG and fMRI did not vary significantly on the channel selected for the EEG-Scale-free coupling. There has been a significant association in the THA (r=0.6857; p=6.00849 10-4\0.05/24, adjusted by the Bonferroni method). The correlation was high but not as significant for the somatosensory network (SOM). (r = 0.4848; p = 0.0259, respectively). The authors discovered that scale-free activity had a strong temporal pattern that was regulated by awareness level.

Feige et al. investigated the relationship between EEG frequency band and fMRI BOLD throughout the whole brain using background subtraction processing of EEG-fMRI data acquired from the resting state, with temporal delays of approximately 10.5 s [35]. The authors chose 10 poor sleepers ("5 m, 5 f, 45.2 16.9 yr, Body Mass Index: 22.4 2.0 kg/m2, Pittsburg Sleep Quality Index PSQI 10.6 3.1") and 10 good sleepers ("5 m, 5f, 44.9 12.1 yr, BMI 22.5 2.7 kg/m2, PSQI 3.6 2.2") for the research. The authors used a 1.5-T Magnetom Sonata scanner to collect the MRI data. They also have a 32-channel MR-compatible EEG amplifier to record the electrophysiological traces. Data were processed at frequencies ranging from 0.1 to 200 Hz and saved for further processing. The authors conducted ICA on all EEG epochs obtained from each participant. In order to identify ballistocardiography components (BCGs), the authors computed ECG-triggered means and assessed all ICA elements on their effect on the mean signal. By adding two FFT spectra from 512-point window merging, the authors estimate the Logarithmic (base e) spectrum strength for each element and EEG segment. The authors converted the fMRI data into Talairach space, applied a high-pass filter at 1/128 Hz, and normalized the data regionally using a full-width half-maximum (FWHM). They also conducted deconvolution analysis on the EEG-related fMRI time stream. The authors segmented the deconvolution time courses among all threads by vector length and utilized a robust cluster-based variant of the K-means technique. According to posthoc analyses, poor sleepers had a more significant negative association in the right fusiform gyrus.

Table 1 shows the summary of the recent research on the combined neuroimaging-based drowsiness detection and sleep inertia issues.

## 3. Challenges and Future Directions

The Individual EEG, fNIRS, and fMRI data properties vary widely. Particularly, drowsiness mechanisms differ amongst people. So, a generalized DDD method should be created to overcome inter-individual disparities. Generalized features and DM models research is sparse, as is the generalizability of drowsiness detection techniques over ground truths. With a generalized approach, our team determined drowsiness. They advocated employing a high sample size to ensure maximal stability and interindividual generalizability. Second, distinct cognitive tasks should be appropriate across subjects.

With dry sensors, the EEG signal is weak and susceptible to motion disturbances, the same goes for fNIRS and fMRI. An algorithm that can extract characteristics from motion artifacts is needed. Deterministic models with continuous output are utilized to improve sleepiness detection resolution and early detection, however, a lack of ground truth limits their validation. There are now just discrete output ground facts.

A driver-friendly and efficient EOG-based method would be more attractive to the industry than an EEG-fNIRS and EEG-fMRI-based system due to the real-time capability issue. A commercial EEG-based DDD solution is currently available for the professional drivers of coal mines in Australia, thanks to recent developments in EEG dry sensors, low-power integrated circuits, and wireless communication technology. The same method should be used to enable commercial EGG-fNIRS and EEG-fMRI-based drowsiness detection systems with low cost.

On the other hand, a closed-loop drowsiness detection system relies on the causal relationship between EEG biomarkers and driver drowsiness levels. This causal relationship may be verified by directly manipulating open-loop EEG biomarkers using neuromodulation techniques like tDCS and tACS. Those EEG traits were linked to tiredness in drivers, but we couldn't identify any further research that properly verified the causal association. More research should be done on closed loop drowsiness detection system by combining different neuroimaging techniques.

Table 1

A summary of the reviewed combined neuroimaging-based drowsiness detection papers

| References | Algorithm Applied | Contribution |
|---|---|---|
| [20] | **Software**: ISAE-SUPAERO flight Simulator, Matlab R2015b<br>**Hardware**: DR400 aircraft<br>**Algorithm/Methods**: Automatic Subspace Reconstruction, butterworth filter, Modified Beer-Lambert Law | • Classify the variation of cognitive<br>• Success Rate 87.2% (Simulation Software)<br>• Success Rate 87.6% (Actual flight Condition) |
| [21] | **Software:** MATLAB, open SIFT, BCILAB, OpenMEEG, MoBILAB toolbox<br>**Hardware:** 64 – channel wireless dry headset, Cognionics hardware, padded sensor, dry electrode,<br>**Algorithm/Methods:** Boundary Element Method, Bayesian model averaging in EEG/MEG imaging, SVD-based reformulation | • Proposed system is capable of real-time analysis<br>• Proposed system differentiates among simulated data and real EEG data obtained from a unique wearable high density (64-channel) dry EEG system |
| [22] | **Software:** EEGLAb toolbox, MATLAB, nirsLAB toolbox, SPSS<br>**Hardware:** 32-bit Analog to a Digital Converter, Synamp2, NIRScout, electroencephalographical sensors, 3D-digitizer.<br>**Algorithm/Methods:** FFT, high pass filter, low pass filter, the wilcoxon signed-rank test, | • Proposed lane-departure scenario with concurrent EEG-fNIRS recordings<br>• Measured EEG-fNIRS responses to lane-departure driving tasks |
| [24] | **Software:** WTK software, MATLAB, EEGLAB Toolbox, nirsLAB toolbox,<br>**Hardware:** V-Amp 16 expert system, NIRScout, NIRS cap,<br>**Algorithm/Methods:** FFT, modified Beer-Lambert law | • Using simultaneous fNIRS and EEG signals to monitor the relationship between hemodynamics and electrical functions and driving performance<br>• Found a greater percentage of oxygenated hemoglobin (HbO2) and EEG theta, alpha, and beta changes |
| [25] | **Software:** BCI2000 software, EEGLAB, MATLAB,<br>**Hardware:** 64 electrodes with custom-built fNIRS system, webcam<br>**Algorithm/Methods:** Fisher's linear discriminant analysis (FLDA), Beer-Lambert's law (mBLL) | • disclosed the neuro-physical relation of driver fatigue through simultaneous EEG, ECG and fNIRS data<br>• Accuracy rate 68.3% in detecting drowsiness. |
| [26] | **Software:** Notch filter, Bandpass filters, Gran Turismo, Biosemi Active Two system, BCI 2000,<br>**Hardware:** Driving simulator system, HD Pro Web-Cam C920, Biosemi Active Two system, fNIRS system<br>**Algorithms/Methods:** FLDA based classifier | • presented a combined EEG and NIRS method to indicate an awake to drowsy state transition<br>• Accuracy Rate 79.2% of detecting drowsiness |
| [27] | **Software:** Brain Vision Analyzer 2.1, BrainVision RecView, CONN toolbox v17f, Artifact Detection Toolbox, ezANOVA,<br>**Hardware:** MR-compatible cap, Magnetom Prisma 3T scanner, wrist actimeter, EyeLink 1000 fMRI eye-tracking system<br>**Algorithm/methods:** Semi-automatic MRI artifact rejection, false discovery rate (FDR), cole-Kripke algorithm, Tudor-Locke algorithm, Benjamini–Hochberg | • brain changes in behavior execution, EEG spectral power, and fMRI functional resting-state connection<br>• Found impaired performance at the DST at awakening and an intrusion of sleep-specific features into wakefulness brain activity |
| [29] | **Software:** MATLAB 2008b, SPM8<br>**Hardware:** MR-compatible EEG device, eight-channel angiography, 1.5-T scanner, MR scanner, MR-compatible electrostatic head phones.<br>**Algorithm/Methods:** conventional visual scoring, Valley technique, Multivariate analysis of variance | • Defined the mechanism by which task-related and task-negative brain activity adjusts in TSD<br>• Used a multimodal method, including concurrent fMRI and EEG, to identify brain regions activated to sustain attentional control |
| [32] | **Software:** GIFT toolkit, SPM12<br>**Hardware:** BrainAmp MR plus, SyncBox MainUnit, 3 T Siemens Trio scanner, fMRI scans, MRI-compatible EEG cap<br>**Algorithm/Methods:** ChebyshevII-type filter, FFT, PCA, standard template subtraction | • Identified dynamic brain network variations from multi-modalities standpoint in sleep inertia.<br>• Found drowsy characteristics of decreasing EEG power and reduced BOLD function when an individual is in sleep inertia |
| [34] | **Software:** EPI sequence, EEGLAB FMRIB toolbox,<br>**Hardware:** Non-magnetic MRI-compatible EEG device, 3T Siemens Trio scanner<br>**Algorithm/Methods:** ANOVA, post-hoc t-tests, Bonferroni technique | • Proposed combined EEG-fMRI information in healthy persons during awake and NREM sleep<br>• Discovered scale-free activity had a strong temporal pattern that was regulated by awareness level. |
| [35] | **Software:** N/A<br>**Hardware:** 1.5-T Magnetom Sonata scanner, 32-channel MR compatible EEG amplifier<br>**Algorithm/Methods:** FFT, high-pass filter, FWHM | • Discovered relationship between EEG frequency band and fMRI BOLD throughout the whole brain using background subtraction processing of EEG-fMRI data |

Moreover, there are very little research was done combining neuroimaging techniques to detect drowsiness detection and observe brain activity during fatigue and sleep inertia problems. More research should be done to explore the application of the combined neuroimaging techniques in the real-time system. ECG with combined neuroimaging can also be explored to extract more information about brain activity along with heart activity during the drowsy state.

## 4. Conclusion

We make three contributions with this review paper: (1) a comprehensive overview, concise description of the BCI based drowsiness detection; (2) overview of the combined neuroimaging-based drowsiness detection system such as EEG+fNIRS, EEG+fMRI; (3) a discussion of various potential ways to improve the existing combined neuroimaging-based drowsiness detection system. The combined neuroimaging techniques for drowsiness detection system is still in the development phase, more research should be done prior to commercial use.

## Acknowledgments

This research did not receive any external funds.

## Conflict of interest

The authors declare that there is no conflict of interest in this paper.